\documentstyle[12pt]{article}

\begin{document}

\centerline{\Large S.P.Novikov}

\vspace{0.3cm}

\centerline{\bf On the exotic De-Rham cohomology.}
 \centerline{\bf
Perturbation theory as a spectral sequence.}

 \vspace{0.3cm}

{\Large 1.Introduction. Definitions and Problems.}

 \vspace{0.2cm}

We are going to consider in this work some exotic homological
constructions within the standard De-Rham complex of differential
forms on the closed compact $C^{\infty}$-manifold $M^n$ with
Riemannian metric $g_{ij}$ started by the present author in 1986
(see below). In all cases we consider an operator
$$d_{\lambda\omega}=d+\lambda\omega:\Lambda^*\rightarrow\Lambda^*$$
where $\omega$ is an one-form on the manifold $M^n$, $\Lambda^k$
is a space of all $C^{\infty}$ k-forms on the same manifold,
$\Lambda^*=\sum_{k\geq  0}\Lambda^k$ and
 $$d_{\lambda\omega}(a)=da+\lambda\omega\wedge a; a\in\Lambda^*$$
(the product is always external here). {\bf A form $\omega$ is not
necessarily closed}. As usually, using Riemannian metric we define
an adjoint operator
$$d^+_{\lambda\omega}=\pm*d_{\lambda\omega}*=d^++\lambda\omega^+;\lambda\in
R $$ for the real values of parameter, and a continuation of this
family of   operators to the complex domain
$$d_{\lambda\omega}+d^+_{\lambda\omega}=d+d^++\lambda(\omega
+\omega^+);\lambda\in C$$ This family is elliptic. Let us consider
a $\lambda$-dependent kernel $K(\lambda\omega)$ of the operator
$$\Delta_{\lambda\omega}(a)=(d_{\lambda\omega}+d^+_{\lambda\omega})^2(a)=0$$
in the space of all forms $\Lambda^*$. This kernel is always a
$Z_2$-graded subspace

$$K=K^++K^-,K^{\pm}\subset \Lambda^{\pm}$$ where $\Lambda^{\pm}$
are the subspaces of the forms of the even (odd) dimension. For
the closed form, $d\omega=0$, these kernels are always $Z$-graded
$$K=\sum_{k\geq 0}K^k$$
\newtheorem{df}{Definition}
\begin{df}
We call the dimension $b^{\pm}(\lambda\omega)$ of the kernel
spaces $K^{\pm}(\lambda\omega)$ by the {\bf (metric dependent)
exotic Betti numbers of the first kind} for the Riemannian
manifold $M^n$ and the 1-form $\omega$. For the closed form
$d\omega=0$ they are $Z$-graded
$$\sum_{k=2l}b_k(\lambda\omega)=b^+;\sum_{k=2l+1}b_k=b^-$$
 and
homotopy invariant.
\end{df}

They can be presented as homology and cohomology groups of the
following complex presented in the form of the short exact
sequence $$0\rightarrow \Lambda^+(M^n)\rightarrow
\Lambda^-(M^n)\rightarrow 0$$ with differential
$D=d_{\lambda\omega}+d^*_{\lambda\omega})$. Its adjoint complex
has a form $$0\rightarrow\Lambda^-\rightarrow\Lambda^+\rightarrow
0$$ with the same operator $D^*$. This homological treatment will
be used in the paragraph 2 for the construction of perturbation
theory as a spectral sequence of such complexes.

 At the same time we define also another homological construction
based on the metric independent complex
$\Lambda^*_{\Omega},d_{\lambda\omega}$ where $\Omega=d\omega$ and
$a\in \Lambda^*_{\Omega}$ if and only if

 $$0=\Omega \wedge a=d_{\lambda\omega}^2
 a$$
The space $\Lambda_{\Omega}$ is independent on $\lambda$. Its
definition is purely local, in the tangent space of every point.
\begin{df}
We call the homology groups $H^k_{\lambda\omega}$ of the complex
$\Lambda^*_{\Omega},d_{\lambda\omega}$ by the {\bf (metric
independent) exotic homology groups of the second kind}. Their
dimensions $b' _{\lambda\omega}$ (if they are finite) we call
exotic Betty numbers of the second kind. For the closed 1-form
$\omega$ they are homotopy invariant and
$\Lambda^*_{\Omega}=\Lambda^*(M^n);\Omega=d\omega=0$ .
\end{df}

 This is a partial case of the very general construction: for any
 space $\Lambda$ and operator
$A:\Lambda\rightarrow \Lambda$ we always can define homology group
for the subspace $\Lambda_A=Ker A^2$ with operator $ A$ as a
differential $$H(\Lambda_A,A)=Ker A/A(Ker A^2)$$

 Following simple lemma has been used
 by the present author and A.Pazitnov in 1986 (see \cite{N,P}; it
looks like this elementary important statement never has been
mentioned before that in the literature):
\newtheorem{lm}{Lemma}
\begin{lm}.1)
For the closed one-form $d\omega=0$ the first and the  second kind
exotic homology groups exactly  coincide with each other and with
the homology groups $H^*_{\rho}(M^n,C)$ with values in $C$ where
the representation of fundamental group
$\rho:\pi_1(M^n)\rightarrow C^*$ is given by the formula
 $$\rho(\gamma)=\exp\{\int_{\gamma}\lambda\omega\}$$
2)For any pair of the (possibly nonclosed) one-forms
$\omega,\omega'$ such that $\omega-\omega'=df,f\in
C^{\infty}(M^n)$, these homology groups are isomorphic to each
other. On the level of forms this isomorphis is given by the gauge
transformation $$a\rightarrow e^{\lambda f}a;d\rightarrow
e^{\lambda f}de^{-\lambda f}=d+\lambda df\wedge$$
\end{lm}
Therefore, for the given closed form $\omega$ we have a complex
curve $\rho\in \Gamma_{\omega}$ in the representation space
$(C^*)^p,p=b_1=dim [H_1(M^n,R)]$.Let us denote by the
$\gamma_j,j=1,2,\ldots p$, the basis of integral cycles in the
group $H_1(M^n,Z)/Torsion$. Our $\lambda$-dependent complex curve
$\Gamma_{\omega}$ has a form: $$z_j=w^{\alpha_j}$$
 where
$$z_j=\rho(\gamma_j);
w=e^{\lambda};\alpha_j=\oint_{\gamma_j}\omega$$

 As a corollary from the lemma, we are coming to the
following

{\bf Conclusion.} The complex curve $\Gamma_{\omega}$ in the space
$(C^*)^p\subset CP^p$ is algebraic  if and only the cohomology
class $[\lambda_0\omega]$ is quantized for some value
$\lambda_0\neq 0$, i.e. it belongs to the integral part
$H^1(M^n,Z)\subset H^1(M^n,R)$.

We started in 1986 to study the following problem (see \cite{NS},
Appendix 1):

{\bf Problem 1}. What are the ''right'' Morse-type analytical
estimates for the numbers of critical points $x_l\in M^n$ for the
one-form $\omega=(\omega_i)$ or for the vector field
$\eta=(\eta^i)=\omega_jg^{ij}$ where $\omega(x_l)=0$ or
$\eta(x_l)=0$? Are these numbers connected with such  analytical
invariants of the one-form $\omega$  like the exotic Betti numbers
of the first and second kind defined above?

Everybody knows   classical Morse-Smale inequalities for the exact
one-forms $df=\omega$ (i.e. for the one-valued real generic
functions $f$ on the manifold) found by the methods of the
classical  topology. Witten (\cite{W}) found a beautiful
analytical treatment of this problem in 1982 based on the
semiclassical zero modes for the operators
 $\Delta_{\lambda\omega}=(d_{\lambda\omega}+d^*_{\lambda\omega})^2$
where $\lambda\in R$ is real, $\lambda\rightarrow \infty$ and
$\omega=df$.

 For the closed one-forms the
proper analog of the Morse-Smale inequalities was found by the
present author in 1981 as a part of the ''multivalued calculus of
variations'' (see \cite{N1}) using differential topology and
homology with coefficients in some specific (''Novikov'') rings.
Their sharpness was proved in the work \cite{F} for the manifolds
with $\pi_1=Z$.  This approach was developed in the works
\cite{F,P,P1,P2,Le} and by others later (let us mention that in
the work \cite{P1,P2} an important lemma of Sikorav has been used
who never published it). The application of Witten analytical
method to this problem was started by Pazitnov in \cite{P}.

The problem is especially interesting for the generic vector
fields $\eta$ or for the non-closed one-forms
$\omega=(\omega_i)=(g_{ij}\eta^j)$: the present author found in
\cite{NS}, Appendix 1, something like the analog of  Morse
inequalities for vector fields based on  the exotic first kind
Betti numbers using an extension of the Witten method. In
particular, a nice diagonalization of the  real fermionic
quadratic forms was found  as a lemma needed for this (see also in
\cite{S} and in a recent popular text \cite{N2}):

$$m_{\pm}(\eta)\geq \max_{g_{ij}}\{ \limsup_{\lambda\rightarrow
\infty} b^{\pm}(\lambda\omega)\}$$

where $\lambda\in R$.
 Here $m_{\pm}(\eta)$ are the numbers of positive (negative) singular
points $\eta=0$ for the generic vector field $\eta$.

{\bf Problem 2}.  How to calculate effectively the numbers in the
right-hand part of this inequality? They are metric independent.
 Is it possible to
express them through the topological characteristics of the
dynamical system (or 1-foliation) determined by the vector field
$\eta$?

The local topological invariants of the critical points of vector
fields are essentially different from the ones for the
differential 1-forms in the absence of metric. The only
topological invariant of the critical point $\eta=0$ for the
vector field is a sign of the determinant $$ \det
\left(\frac{\partial \eta^i}{\partial y^j}\right )(x_k)$$ where
$y^j$ are local coordinates near this critical point.

For the 1-form $\omega$ we have a matrix
 $$a_{ij}=\left (\frac{\partial
\omega_i}{\partial y^j}\right )(x_k)$$
 In the generic case we have
also this invariant but we also have generically nondegenerate
symmetric part of this tensor
 $$\det
[a_{ij}+a_{ji}]\neq 0$$
 Therefore a number of negative squares for
the symmetric part  is also a local topological invariant for this
tensor with two lower indices . We may expect a $Z$-graded theory
for this problem (if we are lucky).

{\bf Question}. What is a geometrical meaning of the metric
independent exotic homology groups of the second kind for the
nonclosed differential 1-form $\omega$? Are they connected with
the critical points of this form? If the answer is  ''yes'',  we
should consider all family of 1-forms $\exp \{f(x)\}\omega$ for
the arbitrary smooth function $f$. How they are connected with
properties of the tangent hyperplane distribution $\omega=0$
especially in the case when this distribution is integrable
$\omega \wedge d\omega=0$? What is happening when we have an
opposite ''contact'' case $\omega\wedge (d\omega)^n\neq 0$? In the
last case how are they depend on the properties of the ''Reeb''
1-foliation determined by the kernel of the form $d\omega=0$?

\pagebreak

\centerline{\Large 2. Closed 1-forms}

\vspace{0.3cm}

This case is the most classical. It essentially was investigated
in 80s as it was already mentioned. The exotic Betti numbers of
the first and second kind coincide with each other. They are
homotopy invariant. Any closed one-form $\omega$ is locally exact
$\omega=df_U$ in any small enough domain $U$. Therefore we have a
local isomorphism of the sheafs of $d$-algebras of differential
forms in any small open domain $U$

$$\Lambda^*_U\rightarrow \Lambda^*_U;d\rightarrow
d+\lambda\omega$$ where
 $$d+\lambda\omega=e^{\lambda
f_U}de^{-\lambda f_U}$$

This local isomorphism leads to the identification of the exotic
homology theories defined by the operator $d_{\lambda\omega}$ with
the $C$-valued homology defined by the family of representation
$\rho(\lambda): \pi_1(M^n)\rightarrow C^*$ such that
$$\rho(\gamma)=\exp\{\oint_{\gamma}\lambda\omega\}$$ for the closed path
$\gamma$. Let $\gamma_i,i=1,2,\ldots p, p=b_1(M^n)$, is some basis
in the first integral homology group modulo torsion, and
$$\alpha_j=\oint_{\gamma_j}\omega$$

 As  already was mentioned in
the introduction, this identification leads to the following
\begin{lm}
The family $\Gamma_{\omega}$ of representations $\rho(\lambda)$
has a form $$z_j=w^{\alpha_j};w=e^{\lambda};z_j=\rho(\gamma_j)$$
In particular, this complex curve in the representation space
$(C^*)^p$ defines an algebraic curve $\Gamma_{\omega}\subset CP^p$
if and only if all ratios $\alpha_j/\alpha_i$ are rational
numbers.
\end{lm}
According to the authors work \cite{N}, there is a finite number
of algebraic subvarieties $W^k_l\subset (C^*)^p\subset CP^p$ in
the space of representations (defined by the polynomials over $Z$)
such that some Betti number $b_k^{\rho}=dim[H^{\rho}_k(M^n;C)]$ is
larger than in the generic point of the representation space
$$b_k^{\rho}>b_k^{min}+l-1,l=1,2,\ldots ;\rho\in \bigcup_l W^k_l$$
 ({\bf jumping
subvarieties}). These polynomials can be considered as a natural
and very general extension of the Alexander-type polynomials.
Their zeroes on the space of representations are exactly the union
of these subvarieties $W=\bigcup W^k_l$ (see the works
\cite{A,Le,La} where the knot and link complementary domains were
studied from that point of view).

 In our  work \cite{N}
these subvarieties were invented for the needs of the Morse theory
for the 1-valued real generic functions: we found out that the
ordinary Morse inequalities can be easily extended to the homology
with representations $$m_k(f)\geq max_{\rho }
b_k^{\rho}(M^n,C^r)/r;\rho:\pi_1(M^n)\rightarrow GL_r(C)$$

 (and
even to the infinite-dimensional von Neuman $II_1$ factors
$r\rightarrow\infty$ in the joint work with Misha Shubin
\cite{NS}).  We looked for the maximal Betti number on the space
of representations
 $$m_k(f)\geq
b_k^{max}/r$$ where $b_k^{max}=max_l[b_k^{min}+l]$ such that
$W_l^k$ is nonempty.

A program of investigations has been proposed by the present
author in the late 80s to apply these ideas to the knot theory
using higher dimensional representation spaces ($r>1$) of
fundamental group of the complementary domain. Very interesting
results were obtained in the work \cite{Le1}.

 Let us return to the case $r=1$. Consider  a closed 1-form $\omega$ with the
Morse critical points whose numbers (with Morse index  $k$) are
equal to $m_k(\omega)$.

 For the special case of the ''rational'' 1-forms
considered by Pazitnov in \cite{P},  the complex curve
$\Gamma_{\omega}$ became algebraic and rational. He worked in fact
only with this special family. All intersections $W_l^k\bigcap
\Gamma_{\omega}$ became zero-dimensional and algebraic. Therefore
this set is  finite in the set of representations. Its inverse
image in the $\lambda$-plane may be infinite but for the
corresponding sequence $\lambda_j\rightarrow\infty$ we have
$Re[\lambda_j]<C$. We cannot use them at all for the Witten-type
method. It certainly needs the property
$Re[\lambda_j]\rightarrow+\infty$.

 Following two statements are true\footnote{ David Hamilton
from the University of Maryland helped me to prove the first of
them. Boris Mityagin from the Ohio State University helped me to
clarify  this situation completely; we removed jointly an
important mistake.}

\newtheorem{tm}{Theorem}

\begin{lm}
For any nontrivial algebraic variety $W$ in the compactified space
of representations $\pi_1\rightarrow C^*$ of the codimension one,
and any irrational closed 1-form $\omega$, the intersection
$\Gamma_{\omega}\bigcap W$ contains infinite number of points
$\lambda_j\in (C^*)^p$ such that $\lambda_j\rightarrow\infty$ and
$|Re[\lambda_j]|<C$.
\end{lm}

\begin{tm}(B.Mityagin, S.Novikov)
For any compact manifold $M$ (or for any finite simplicial complex
$M$)  and for any closed real 1-form $\omega$ the intersection of
the jumping subvariety with the curve determined by the form
$\omega$, $W\bigcap \Gamma_{\omega}$ contains only a finite number
of points $\lambda_j\in D$ in any domain $D$ of the
$\lambda$-plane such that its intersection with every closed strip
parallel to the imaginary axis $a\leq Re[\lambda ] \leq b$, is
compact.
\end{tm}

Proof of these facts is not hard.

 As a
{\bf conclusion}, we see that Witten-type method leads here always
to the same estimate if manifold $W$ does not contain  curve
$\Gamma_{\omega}$. As it follows from the old results  \cite{F,P},
this minimal Betti number in the generic point $\lambda_{generic}$
is equal to the so-called ''Novikov Betti number'' defined in
\cite{N1}. It was written in \cite{F}, Appendix,  in the purely
algebraic  language of modules started in these problems by Milnor
in \cite{M}, for the integral cohomology classes only, and space
of representations was not discussed. In the works \cite{N,P}
everything was written (first time) in the form of analysis on the
space of representations, but only integral cohomology classes
were considered in \cite{P}.

 Let me remind here the calculation of the generic Betti
numbers $b_k^{min}$ on the spaces of one-dimensional
representations based on the {\bf analytical perturbation theory
in the form of spectral sequence} near the point $\lambda=0$ made
by the present author in 1986  in the work \cite{N}. This idea is
very simple and effective:

Starting from the point $\lambda=0$, we consider the De-Rham
complex of the formal series $a(\lambda)=a_0+\lambda a_1+\lambda^2
a_2+\ldots$, with differential $d_{\lambda\omega}=d+\lambda\omega$
as before. The equation $d_{\lambda\omega}a(\lambda)=0$ leads to
the equalities
$$da_0=0;da_1=-a_0\wedge\omega;\ldots,da_s=-a_{s-1}\wedge\omega$$
As a corollary from these arguments, we are coming to the
following result (\cite{N})
\begin{tm} Spectral sequence $E_s,d_s$ is well-defined and homotopy invariant
where $d_s^2=0$
and $E_{s+1}=H(E_s,d_s)$, with differentials defined by the
formula
 $$d_0=d,E_0=\Lambda^*$$
$$E_1=H^*(M^n,R);d_1(a)=[\omega]\wedge a$$
$$d_s(a)=\{\omega,\omega,\ldots,\omega,a\}_s;a\in E_s$$ Here
$\{b,b,\ldots,b,a\}_s$ is a homotopy invariant ''Massey product''
of the order $s$ defined in the ''refined way''  (every time when
we perform the operation $d^{-1}$ over the same elements, we
should take exactly the same results). In this specific case we
have $$\{b,\ldots,b,a\}_s= b\wedge
d^{-1}\{b,\ldots,b,a\}_{s-1};b=\omega$$
 The groups $E_{\infty}$ are
isomorphic to the homology groups in the generic points of the
representation space whose ranks are equal to the numbers
$b_k^{min}$.
\end{tm}
Let me point out that some spectral sequence was discussed already
in \cite{F},Appendix,  in purely algebraic language of modules
but its differentials were not calculated.

For the algebraic manifolds (and in general for the ''formal''
spaces) the Massey products disappear. Only one nontrivial
differential $d_1(a)=\omega\wedge a$ remains. Some people met this
special situation later (in the late 80s and 90s). For the
symplectic manifolds (like nil-manifolds, for example) there are
plenty of nontrivial Massey products. Our spectral sequence will
be nontrivial in these cases (see \cite{A1}). Such arguments can
be used also for the higher dimensional representation spaces
$\rho:\pi_1(M^n)\rightarrow GL_r(C)$ constructing the perturbation
theory near the trivial representation. A homotopy invariant
spectral sequence appears (see\cite{A}). However, it is difficult
to identify the limiting terms of these sequences because  unit
point is a complicated singularity for the higher dimensions
($r>1$).

 In the
next paragraph we are going to apply the same arguments studying
the exotic first kind Betti numbers defined by the non-closed
1-forms.

\pagebreak

\centerline{\Large 2.The exotic Betti numbers of the first kind.}
\centerline{\Large Perturbation theory as a spectral sequence}

\vspace{0.3cm}

As it was established by the present author in the work \cite{NS},
Appendix 1, the semiclassical and fermionic analysis of the
operator $(d_{\lambda\omega}+d^*_{\lambda\omega})^2$  leads to the
{\bf Morse-type estimate for the generic vector field $\eta$}:
$$m_{\pm}(\eta)\geq
\max_{g_{ij};\omega'}\{\limsup_{\lambda\rightarrow\infty}[b_{\pm}(\lambda\omega')]\}
$$ Here $\lambda\in R$ and $\eta^i=g^{ij}\omega_j$ is a generic
vector field, $m_{\pm}$ are the numbers of its singular points
$\eta=0$ with topological sign $\pm$. The form $\omega'$ can be
obtained from $\omega$ by the following operations:
$$1)\omega\rightarrow e^f\omega=\omega'$$

2)$\omega'$ is a small  perturbation of the form $\omega$ such
that all singular points remain topologically unchanged.

 The detailed
exposition of these arguments (including the effective
diagonalization of the real fermionic quadratic forms by the real
Bogolyubov transformations) can be found also in later works
\cite{S,N2}). It is interesting to point out that our arguments
lead to the metric independent result.

Let us look more carefully on the right-hand side of this
inequality. The numbers
$\limsup_{\lambda\rightarrow\infty}[b_{\pm}(\omega)],\lambda\in
R,$ for the fixed Riemannian metric  exactly coincide with the
exotic first kind Betti number in the generic point
$b_{\pm}^{min}(\omega)$ if and only if the set of ''jumping'' real
values $\lambda_j\in C$ is bounded. It simply means that there is
a finite number of real points such that
$b_{\pm}(\lambda_j\omega)>b_{\pm}^{min}(\omega)$. Maybe also some
specific sequences of the complex points can be used where
$Re[\lambda^2_j]\rightarrow +\infty$. If this set is unbounded, we
may have numbers in the right-hand side larger than the generic
Betti number $b_{\pm}^{min}(\omega)$. For the closed irrational
1-forms $\omega$ we may have an unbounded set of the jumping
points $\lambda_j$ as it was demonstrated in the previous
paragraph but they are ''almost imaginary'' and cannot be used in
the Witten-type method. So it is difficult to calculate these
numbers in general. Even if they are exactly equal to the numbers
$b_{\pm}^{min}$ for some Riemannian metric, we need to find their
maximal value for all metrics in order to compute the right-hand
side numbers.

{\bf Question:} Are the numbers
$\limsup_{\lambda\rightarrow\infty}[b_{\pm}(\omega)], \lambda\in
R$,  equal to $b_{\pm}^{min}$  for the generic Riemannian metrics
and 1-forms? Can they  be larger for the exceptional metrics?

Indeed, Dan Burghelea informed me very recently that he can prove
metric independence of the generic numbers $b_{\pm}^{min}(\omega)$
but later he withdrew his announcement.

{\bf How to calculate the numbers $b_{\pm}^{min}(\omega)$? }

In order to solve this problem, we develop a perturbation theory
for the zero modes near the exceptional value $\lambda=0$.
Consider the equation
$$[d_{\lambda\omega}+d^*_{\lambda\omega}]a(\lambda)=0$$
 where
$a(\lambda)$ is a formal series in the variable $\lambda$
$$a(\lambda)=a_0+a_1\lambda+\ldots +a_s\lambda^s+\dots$$

As before, we are coming to the equalities: $$[d+d^+]a_0=0$$
$$[d+d^+]a_1=-[\omega+\omega^+]a_0$$

$$............................$$
$$[d+d^+]a_s=-[\omega+\omega^+]a_{s-1}$$ for all positive integers
$s\rightarrow\infty$. Here $\omega$ means a multiplication
operator $a\rightarrow \omega\wedge a$, $A^+$ always mean an
adjoint operator for the   operators $A$. For the operators
$d_{\lambda\omega}$ we define
$d^*_{\lambda\omega}=d^+_{\lambda\omega}$ for the real $\lambda$
(the form $\omega$ is also real here). The continuation to the
complex values of $\lambda$ is assumed to be holomorphic as
before.

Let us denote the operators $d+d^+$ by $A$ and $\omega+\omega^+$
by $B$. We denote also by $P_h$ a projection operator on the
subspace of harmonic forms in the given Riemannian metric,
$$P_h^2=P_h:\Lambda^*\rightarrow \Lambda^*$$
\begin{lm}Let for the harmonic form $a_0$ the forms $a_s$ can be defined
by the formula $$a_s=(-1)^s(A^{-1}B)(A^{-1}B)\ldots
(A^{-1}B)(a_0)$$ for all positive integers $s\geq 1$. (The
operator $(A^{-1}B)$ is applied $s$ times in this formula. This
operator is not everywhere defined; it is  multivalued.) Than
there exists positive $\epsilon>0$ such that for all small
$\lambda$ in the domain $|\lambda|<\epsilon$ a $\lambda$-family
$b(\lambda)$ of the zero modes
$$[d_{\lambda\omega}+d^*_{\lambda\omega}]^2b(\lambda)=0$$ with the
initial value  $b(0)=a_0$ is well-defined.
\end{lm}

This lemma on the formal level follows immediately from the
expression above for the coefficients $a_s$. On the analytical
level its proof requires additional analytical arguments. However,
they are not complicated.

>From this lemma we can extract the {\bf adjoint pair
$E^{\pm}_s,D^{\pm}_s$ of spectral sequences of the perturbation
theory}. The idea to construct the differentials $D^{\pm}_s$ is
following:

By definition, the zero term $(E^+_0,D^+_0)$ looks like  the
short exact sequence  below $$0\rightarrow
\Lambda^+\rightarrow\Lambda^-\rightarrow 0$$ where $\Lambda^{\pm}$
is the space of $C^{\infty}$ differential forms on the manifold
$M^n$ of the even (odd) dimensions. The differential $D^+_0$ is
equal by definition to the operator $A=d+d^+$. The homology groups
of this operator are represented exactly by the harmonic forms
$h^+\subset\Lambda^+$ and $h^-\subset\Lambda^-$.

 In order to
construct the operator $D^+_1$ acting on the harmonic forms
$$D^+_1:h^+\rightarrow h^-$$
 let us point out that there is
an obstruction for  solving the equation
$$[d+d^*]a_1=-[\omega+\omega^+]a_0$$
 The harmonic projector $P_h$
annihilates the left-hand part $P_hA=0$ or $$P_h[d+d^+]a_1=0$$
Therefore we need to have $P_hB(a_0)=0$ or
$$P_h[\omega+\omega^+]a_0=0$$ in the right-hand side. We define an
operator $D^+_1$ by the formula $$D^+_1(a_0)=P_hB(a_0)$$

So we have finally a term $E^+_1,D^+_1$ presented in the form of
short exact sequence $$0\rightarrow h^+\rightarrow h^-\rightarrow
0$$

>From that we have a second term $E^+_2$ consisting of  the spaces
$Ker D^+_1\subset h^+$ and $Coker D^+_1\subset h^-$. Let us now
construct $D^+_2:Ker D^+_1\rightarrow Coker D^+_1$ in order to
construct a pair $E^+_2,D^+_2$. We define it by the formula $$
D^+_2(a_0)=P_hB\{u+A^{-1}B(a_0)\};u\in h^+$$ where $u$ appears as
a result of the indefinicy in the inversion of the operator $A$.
Our definition exactly defines a map $$D^+_2:Ker D^+_1\rightarrow
Coker D^+_1$$

Continuing this process, we are coming to the spectral sequence
$E^+_s,D^+_s$ in the form of  short exact sequences $$0\rightarrow
E^{++}_s\rightarrow E^{+-}_s\rightarrow 0$$ with differentials
$D^+_s$ where $E^{++}_1=h^+,E^{+-}_1=h^-$.

In the same way we construct a second (adjoint) spectral sequence
$E^-_s,D^-_s$ in the form of  short exact sequences $$0\rightarrow
E^{-+}_s\rightarrow E^{--}_s\rightarrow 0$$ such that
$E^{--}_1=h^-$ and $E^{-+}_1=h^+$. Finally we are coming to the
theorem summarizing the {\bf Perturbation Theory}:

\begin{tm}
Two spectral sequences  $E^{\pm}_s,D^{\pm}_s$  in the form of
series of the short exact sequences $$0\rightarrow
E_s^{\pm,+}\rightarrow E^{\pm,-}_s\rightarrow 0$$ are well-defined
and adjoint to each other corresponding to the standard inner
product on the spaces of differential forms generated by metric.
Their first terms $E^{\pm,+}_1$ are isomorphic to the spaces of
harmonic forms $h^{\pm}$;  the spaces $E^{\pm,-}_1$ are isomorphic
to the spaces $h^{\mp}$ . Their infinite terms $E^{++}_{\infty}$
and $E^{--}_{\infty}$ have dimensions exactly equal to the generic
first kind Betti numbers $b_{\pm}^{min}(\omega)$. The
differentials in these spectral sequences have a form
$$D^{\pm}_s(a_0)=P_hA^{-1}B\ldots A^{-1}B(a_0)$$ where $A=d+d^+$
and $B=\omega+\omega^+$ for the harmonic form $a_0$.
\end{tm}

As B.Mityagin pointed out to the author, the convergency
 can be easily proved because there is a gap in the
spectrum of the operator $\Delta$ near the zero. It is  standard
exercise in the classical functional analysis where  these kind of
arguments were frequently in use since the times of John von
Neumann. Indeed, different  geometers beginning from the classical
works of Kodaira and Spencer about 40-50 years ago considered
something like spectral sequences of the perturbation theory for
different problems involving deformations. The present author
constructed and calculated through the iterated Massey products
this spectral sequence for the elliptic complex based on the
family of operators $d+\lambda\omega$ where $d\omega=0$, in the
work \cite{N} in 1986. Later, in the
  work \cite{F1}, theorem 6.1, this analytically-based
spectral sequence of perturbation theory was extended to the
family of general elliptic complexes.

 Our goal here is
to study this spectral sequence effectively for the same operators
as in the work \cite{N} but based on the nonclosed 1-forms
$\omega$. In the point $\lambda=0$ their initial term has the same
homological form as before but for the nonclosed forms we are
coming to the fundamental

{\bf Question.} How these spectral sequence  depend on metric for
the nonclosed forms ? We already know that  dimension of the first
term is metric independent. Probably it is true also for the
infinite term. What about intermediate terms? Do their dimensions
depend on metric? What is happening for the generic metric?

The Poincare duality operator $*$ commutes with all differentials
of the spectral sequences.

\pagebreak

\centerline{\Large Appendix. Some remarks} \centerline{\Large on
the exotic second kind De-Rham cohomology groups}

\vspace{0.3cm}

Our general construction of the second kind exotic De-Rham
cohomology groups $H^k_{\omega}$ is following:  take 1-form
$\omega$ and consider a complex $\Lambda^*_{\Omega},d_{\omega}$
where $a\in\Lambda_{\Omega}^*$        if and only if
$a\wedge\Omega=0;\Omega=d\omega$. We denote the cohomology groups
of that complex by $H^*_{\omega }(M^n,R)$. They can be infinite
dimensional for the compact manifolds sometimes.

As before, we can consider a family of operators
$d_{\lambda\omega}$ and construct perturbation theory near the
point $\lambda=0$ for the calculation of these groups in the
generic point $\lambda\neq 0$. We are coming to the spectral
sequence $E_s,d_s$ such that $E_0,d_0=\Lambda^*_{\Omega},d$, and
$d_s(a)=(\omega,\ldots,\omega,a)$ is a Massey product where
$d_1(a)=\omega\wedge a$ is an ordinary product. These operations
are well defined in this complex because $\Omega=d\omega$ and
$\Omega\wedge a=0$ but the form $\omega$ is not a cocycle in this
complex except the ''completely integrable case'' $\omega\wedge
d\omega=0$.

{\bf What geometrical meaning do these homology have?}. We know
one case where this cohomology group has direct meaningful
interpretation: Let the form $\alpha$ determines a completely
integrable distribution $\alpha=0$. By the Frobenius theorem, we
have $d\alpha=\alpha\wedge \omega$. Consider now a first exotic
homology $H^1_{\lambda\omega}$ for the operator $d+\lambda\omega$.
{\bf For the value $\lambda=-1$ we have a nontrivial
one-dimensional exotic cocycle (cohomology class) $\alpha\in
H^1_{-\omega}$ if and only if it defines an integrable
distribution. So the point $\lambda=-1$ is a jumping point, and
corresponding exotic cocycles correspond to the integrable
foliations.}

Consider now some other examples.

Trivial example: For two-dimensional manifolds $M^2$ everything is
trivial: we have $\Lambda^0_{\omega}=0$ and
$\Lambda^*_{\omega}=\Lambda^1+\Lambda^2$. Therefore
one-dimensional exotic homology group is infinite dimensional. It
is equal to the kernel space of the operator $d+\lambda\omega$.
All 1-dimensional distributions are integrable. The second group
$H^2_{\lambda\omega}$ is either one-dimensional (for $\lambda=0$)
or equal to zero (in the generic point).

In the case of three-dimensional manifolds $n=3$ we have two
interesting possibilities except the one discussed above.

I.Let $\omega\wedge d\omega=0$. In any  point $x\in M^3$  with
$rk[\Omega]=2$ there is a local coordinate system $p,q,r$ such
that $\Omega=dp\wedge dq$ and $\omega=pdq$. The equation
$\Omega=0$ determines a ''vortex vector field'' $\eta$ (more
exactly, this is an one-foliation). The equation $\omega=0$
determines a two-foliation. In the local coordinates (p,q,r)
indicated above the foliation is $p=const$; The vortex lines are
given locally by the equation $q=const,p=const$, i.e. it is
tangent to the 2-foliation. The coordinate $r$ is varying along
the ''vortex lines''. Our space $\Lambda^*_{\Omega}$ is isomorphic
to the direct sum $$\Lambda^1_{\Omega}+\Lambda^2+\Lambda^3$$ with
the boundary  operator $d+\lambda\omega$. In the local coordinates
we have $a\in\Lambda^1_{\Omega}$ if and only if it can be written
in the form $$a=fdp+gdq$$ where $f,g$ are the arbitrary functions
of three variables. We obviously have $a\wedge\Omega=0$.

For $\lambda=0$ the condition $da=0$ leads to property that the
closed 1-form $a$ is in fact independent on the variable $r$
locally along the vortex lines. If factor-space by the vortex
lines globally exists as a manifold (or maybe an orbifold), we
have one-dimensional exotic homology infinite-dimensional, coming
from the closed forms in the factor-space. For $\lambda\neq 0$ the
conclusion is the same. All exotic 1-cohomology classes are
represented by the (locally) $r$-independent 1-forms (i.e. coming
from the 'locally defined' factor-space) closed relative to the
$d+\lambda\omega$ operator. We are coming to the conclusion that
the space $H^1_{\lambda\omega}$ is finite-dimensional (and mostly
equal to zero) if the vortex dynamical system is topologically
complicated.

The spaces $\Lambda^k_{**}$ for $k=2,3$ are coincide with the
spaces of all forms. For any $\lambda$ and $k=2,3$ our exotic
homology are locally trivial for the small domains such that $rk[
d\omega]=2 $. Our conjecture is that for the nonsingular
foliations $\omega=0$ these spaces are finite-dimensional.

II. For the contact case we have everywhere $rk[\Omega]=2$, and in
the proper local coordinates $p,q,r$ we have always
$\omega=pdq+dr$,  $\Omega=d\omega=dp\wedge dq$. The spaces of
forms $\Lambda^k_{**}$ for $k=1,2,3$, where the exotic complex is
defined is same here as in the previous example, but the
differential is different. For the 1-form $a$ such that $a\wedge
\Omega=0$ and $(d+\omega)a=0$ we have locally $a=e^ra'$ where $a'$
does not depend on the variable $r$. This property gives us a
(local) sheaf isomorphism between this and previous case. But the
form $dr$ is not well-defined here globally as a closed form. It
simply means that these coefficients has exponential decay along
the vortex lines. Probably in the ergodic case this form should be
equal to zero.

A nice class of example of the  contact manifold where we also
have integrable foliations is class of the (constant energy) phase
spaces $M^3$ of geodesic flows for 2-manifolds of the constant
negative curvature with well known Anosov foliations.

\end{document}